\documentclass{iau}
\usepackage{graphicx}
\usepackage{hyperref}
\title[From Sky to Earth: Data Science Methodology Transfer] %
{From Sky to Earth:\\ Data Science Methodology Transfer}

\author[Ashish A. Mahabal,  D. Crichton, S. G. Djorgovski, E. Law, \& John S. Hughes]   %
{Ashish A. Mahabal$^1$, 
Daniel Crichton$^2$, S. G. Djorgovski$^1$, 
Emily Law$^2$, John S. Hughes$^2$}

\affiliation{$^1$Present address: Center for Data Driven Discovery, Caltech, Pasadena, CA 91125 \\ email: {\tt aam@astro.caltech.edu, djorgovski@cd3.caltech.edu} \\[\affilskip]
$^2$Center for Data Science and Technology, Jet Propulsion Laboratory, Pasadena, CA, 91109 \\
email: {\tt daniel.j.crichton@jpl.nasa.gov, emily.s.law@jpl.nasa.gov, john.s.hughes@jpl.nasa.gov}}

\pubyear{2017}
\volume{325}  %
\pagerange{119--126}
\jname{Astroinformatics}
\editors{M. Brescia, S. G. Djorgovski, E. Feigelson, G. Longo \& S. Cavuoti eds.}
\begin{document}

\maketitle

\begin{abstract}
We describe here the parallels in astronomy and earth science datasets, their analyses, and the opportunities for methodology transfer from astroinformatics to geoinformatics. Using example of hydrology, we emphasize how meta-data and ontologies are crucial in such an undertaking. Using the infrastructure being designed for EarthCube - the Virtual Observatory for the earth sciences - we discuss essential steps for better transfer of tools and techniques in the future e.g. domain adaptation. Finally we point out that it is never a one-way process and there is enough for astroinformatics to learn from geoinformatics as well.

\keywords{Data Science, Machine Learning, Meta-data, Ontologies}
\end{abstract}

\firstsection %
\section{Introduction}
\label{sec:intro}

Datasets in all walks of life are getting bigger as well as more complex owing to increasing number of overlapping non-homogeneous observables, and varied spatial, temporal coverage at different resolutions. As computers become faster, it is the complexity more than the data volume that proves to be a bigger challenge. Different fields like astroinformatics, bioinformatics, geoinformatics are all trying to develop - often with the help of statisticians and mathematicians - tools and techniques to attack problems that are superficially different, but at the core have many similarities. It behooves us to identify such similarities by abstracting the datasets from the raw observations and domain knowledge, and then applying a standardized set of techniques to extract the possible science from the datasets thereby avoiding the reinvention of the proverbial wheel. That is the basic premise behind methodology transfer. A useful side-effect is improved tractability and reproducibility. In practice, the situation is more complex than the brief description above. 

In the following sections we describe the forays in transferring methodology from astronomy and space science to earth science. We start by noting the parallels in these fields (Secn.~\ref{sec:parallels}). We then take the specific case of hydrology, along with the use of ontologies, to illustrate the basic principles of going from data to knowledge (Secn.~\ref{sec:hydrology}). We then describe EarthCube, an equivalent of Astronomy's Virtual Observatory for the Earth (Secn.~\ref{sec:earthcube}). Finally we describe domain adaptation as a specific technique that can be applied to earth data in the same fashion as we are doing in astronomy (Secn.~\ref{sec:da}). This will be  followed by a broader discussion (Secn.~\ref{sec:discussion}).

\section{Parallels in astroinformatics and geoinformatics}
\label{sec:parallels}

\subsection{Astronomy}
\label{subsec:astro}
In the recent past astronomy has been moving from static pictures of small samples of the sky to be analyzed later in the coziness of one's office, to the frantic real-time analysis of movie-like rapid observations of large parts of the sky in pursuit of variations in individual celestial sources that require immediate follow-up. Since, in a given source, time-scales for such a variation to occur would shame glaciers, one has to observe rather large statistically significant populations of objects. Owing to variety in distances, physical conditions, and time-scales involved, a corresponding variety of instruments capable of capturing data at different resolutions, and speeds, is required. The data volumes, currently in tens of TB, are rapidly approaching the PB scales.

To cite specific examples, we have surveys with large number of epochs, and with high-precision, in tiny areas of the sky, like Kepler (\cite{Hall2010}), looking for exoplanets, and surveys like the Catalina Sky Survey (CSS)/Catalina Real-time Transient Survey (CRTS) looking at well over half the sky hundreds of times but at a lower resolution of $\sim2''$ and finding a rich variety of Galactic and extra-galactic variables, as well as asteroids (\cite{Djorgovski2011}, \cite{Mahabal2011}, \cite{Drake2012} etc.).

As we get to know the universe better, we repeatedly find objects of types that we already understand well. Identifying the rare and new phenomena from the complex data is a challenge. Only after these are identified can one follow them up with specialized modes like polarimetry, spectroscopy etc.

Science for a given mission/program is well-defined and carried out by the resident team. But there is always a lot more that is possible. It is when diverse datasets can be combined that these additional goals become reachable with some  ease. The complexity to watch for when using multiple datasets is in terms of:
\begin{itemize}
\item Spatial distribution (data archives at geographically diverse locations),
\item Spatial and temporal resolution of datasets (e.g. Hubble Space Telescope (HST; http://www.stsci.edu/hst/) has an angular resolution of $0''.1$/pixel, and Transiting Exoplanet Survey satellite (TESS) (\cite{Ricker2015}) is proposed to be $10''$/pixel),
\item Number of time epochs and their irregularity (favoring faster or slower phenomena),
\item Overlap in coverage (e.g. DLS (\cite{Becker2004}) covered a small area whereas Gaia (\cite{Brown2016}) is all-sky).
\end{itemize}

In a typical science case scientists have copious data from one survey, and look for cross-matched objects from other surveys, and these data tend to have different resolutions, have been observed at other wavelengths, and may have shorter or longer exposures with smaller or larger aperture telescopes. Good meta-data still allows the combining of such point-wise observations in order to build models involving the variations with time and wavelength. 

\subsection{Earth Science}
\label{subsec:earthsci}

The situation in earth science is analogous to that in astronomy. Many satellites scan the earth at different wavelengths and cadence, resulting in maps that are at different resolutions and with irregular gaps (example observables include moisture, condensation, over- and under-ground water, snowfall etc.). Besides the satellite measurements there are in-situ measurements (e.g. water levels in wells), models (both predictive, and computational). Much like in astronomy these can then be combined in associative or predictive ways. Data volumes often exceed those in astronomy due mainly to the abundance of photons.

\begin{figure}
\begin{center}
 \includegraphics[width=5.0in]{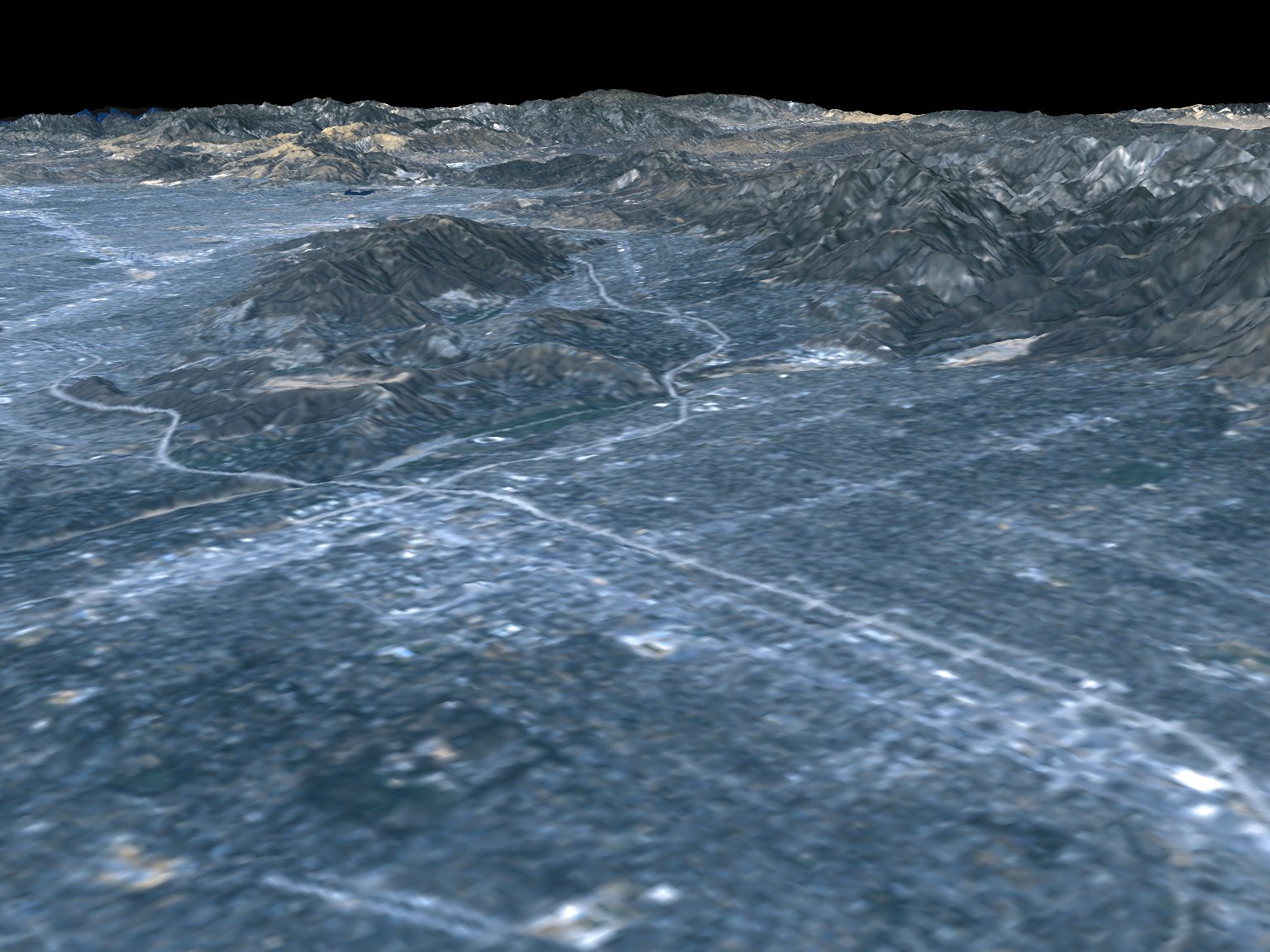} 
 \caption{Perspective view of Pasadena, CA  created by draping a Landsat satellite image (at 30m $\times$ 30m) over a Shuttle Radar Topography Mission (SRTM) elevation model. Topography is exaggerated 1.5 times vertically. The Landsat image was provided by the United States Geological Survey and the image is in the public domain.}
   \label{fig:landsat}
\end{center}
\end{figure}

High spatial resolution from satellites varies from about 30m $\times$ 30m for Landsat (https://landsat.usgs.gov/) images (see Fig.~\ref{fig:landsat} for an example) all the way to 4km $\times$ 4km for Tropical Rainfall Measuring Mission (TRMM; https://pmm.nasa.gov/trmm) radar, which operated for 17 years from 1997 to 2015. The precipitation model of TRMM can be termed as medium resolution with $0^{\rm o}.25 \times 0^{\rm o}.25$, whereas the $1^{\rm o} \times 1^{\rm o}$ resolution of the Atmospheric InfraRed Sounder (AIRS; http://airs.jpl.nasa.gov/) surface air temperature  measurements can be termed as low spatial resolution. The cadence too varies from under one per day, to about two per day. Data latency variability is unique to space-based sensors. It varies anywhere from once every three hours, to once every three months. This is true for astronomical observatories too. TESS, for instance, will downlink once every 15 days. Space-based data tend to be sparser owing to slower links (Gaia downlinks only 1D projections of images), and the computers too tend to be older because space-proofing and extensive testing is required before launch. 

\subsection{Tools}
\label{subsec:tools}

Tools used in both astroinformatics and geoinformatics come in two types: the generic ones include, for example, programming languages that are typically used, as well as libraries based on them. Standardization is more extensive today than years ago, and owing to better connectivity and open source efforts, it is getting even better. Python, for instance, is extensively used, and when advanced statistical methods are required, R is the language of choice. However more established, but licensed, systems like MatLab and IDL also get used (this is true in bioinformatics as well). Things differ when it comes to tools that are not as multi-purpose as languages, and such tools could either be proprietary or open e.g. Geographic Information Software like ArcGIS (https://www.arcgis.com/), Grid Analysis and Display System (GrADS; http://cola.gmu.edu/grads/), HydroDesktop (https://hydrodesktop.codeplex.com/), Google Earth Engine (https://earthengine.google.com/) etc. for earth science. These typically get used for one or more datasets, often voluminous. The difficulty in using them for more datasets is generally the non-uniformity of data-formats. Thus, details at a lower-level get set in stone as the differences get propagated. However, standards from organizations like Open Geospatial Consortium (OGC; http://www.opengeospatial.org/) have gone a long way
to help bring earth science data together. Similarly, the Distributed Active Archive Centers (DAACs)  have contributed to the improvement of interoperability for
earth science data. In astronomy the Flexible Image Transport Format (FITS) (\cite{Pence2010} and references therein) has been used for a long time and that standardization is built-in to almost all datasets, with the flexible format evolving to include data tables, and multiple images, and the standardization has hugely benefited the community.

The standardization effort was greatly boosted by the formation first of the National Virtual Observatory (NVO) (\cite{ Brunner2001}, \cite{Djorgovski2005}) for astronomy, followed by the Virtual Astronomical Observatory (VAO) (\cite{Hanisch2010}). Various  tools were built including the Data Discovery tool\footnote{\url{http://www.usvao.org/index.html\%3Fpage\_id=344.html}} to discover available resources on specific sources, the Cross-matching service\footnote{\url{http://www.usvao.org/index.html\%3Fpage\_id=364.html}} to cross-identify sources in catalogs at different wavelengths, and with different resolutions, Iris, the Spectral Energy Distribution analysis tool (\cite{Laurino2013}, \cite{Laurino2014})
and the Time Series search tool\footnote{\url{http://www.usvao.org/index.html\%3Fpage\_id=370.html}}. Many times some subset standard VO functionality (e.g. basic cross-matching) became so ingrained in the daily scientfic workflow that it often ceased to be recognized as due to the Virtual Observatory. That is the hallmark of a good tool.

The workflow tools used in astroinformatics and geoinformatics are also inherently exchangeable. This is because the progress from measurements, to products, to applications also parallels each other in the two fields. For example, a satellite does regular measurements of the weather, models are then built indicating excess rainfall in a hilly area, and a flash flood warning is issued. Or a telescope observes an area of the sky, new measurements are compared with old observations, a much brighter object is found very close to a  galaxy, and a possible supernova is announced which can then be verified by other teams.

To make the example more concrete, consider this example:
Satellite Aqua using the AIRS sensor measures 3D Atmospheric Temperature, Humidty, Clouds and does such a prediction (ARSET -https://arset.gsfc.nasa.gov/ - has examples). On the other hand, in astronomy,
CRTS uses new CSS observations, compares them with its DR2 catalog, cross-matches with a galaxy catalog, and finds the supernova.

The comparison stops there though. The
data latency for earth observations tends to be more than in astronomy. The various earth data centers are distributed and managed by many different groups that also represent many different sponsors and funding sources.  As such, their level of interoperability it limited.   These generally do not communicate with one another in real-time.  Being distributed is fine, but they should strive to provide interoperability to establish a virtual data environment.  The level of interoperability is often the result of different access methods, data representations, and governance rules. 

The joint
initiative on data science and technology established between JPL and Caltech is focused on improving
methodology transfer by leveraging existing work in astroinformatics and geoinformatics. There exist parallel efforts elsewhere towards this end (for example, in Europe, it is under the aegis of the COST foundation - http://www.cost.eu/). 

\begin{figure}
\begin{center}
 \includegraphics[width=5.0in]{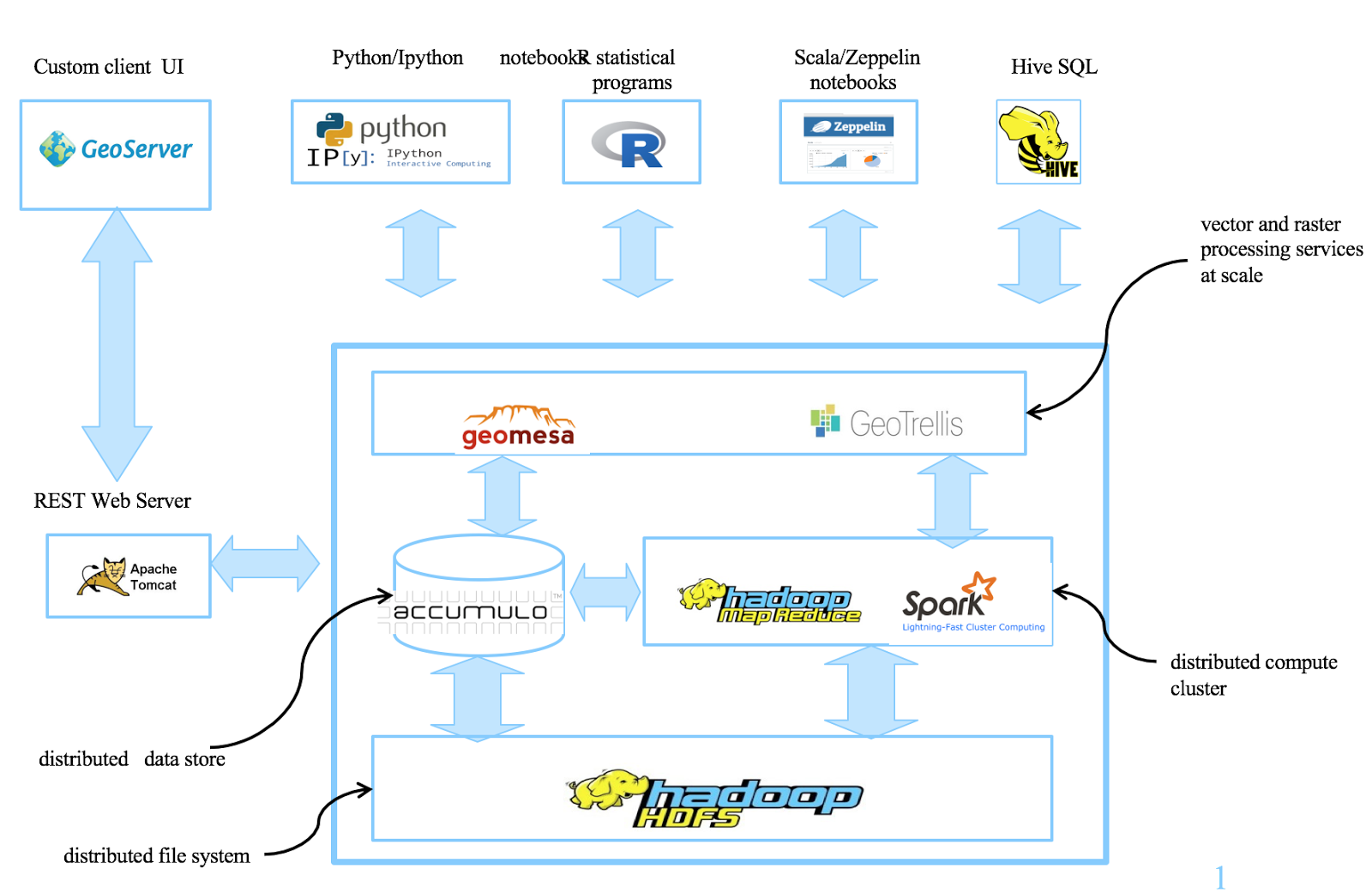} 
 \caption{Architecture for the Water Management stack. Accumulo as the DB, HDFS forming the backup, mapreduce and Spark powering the computation, and GeoMesa/GeoTrellis providing the indexing; various webservices and Python/R providing the interfaces.}
\label{fig:wswmarch}
\end{center}
\end{figure}

\begin{figure}
\begin{center}
 \includegraphics[width=5.0in]{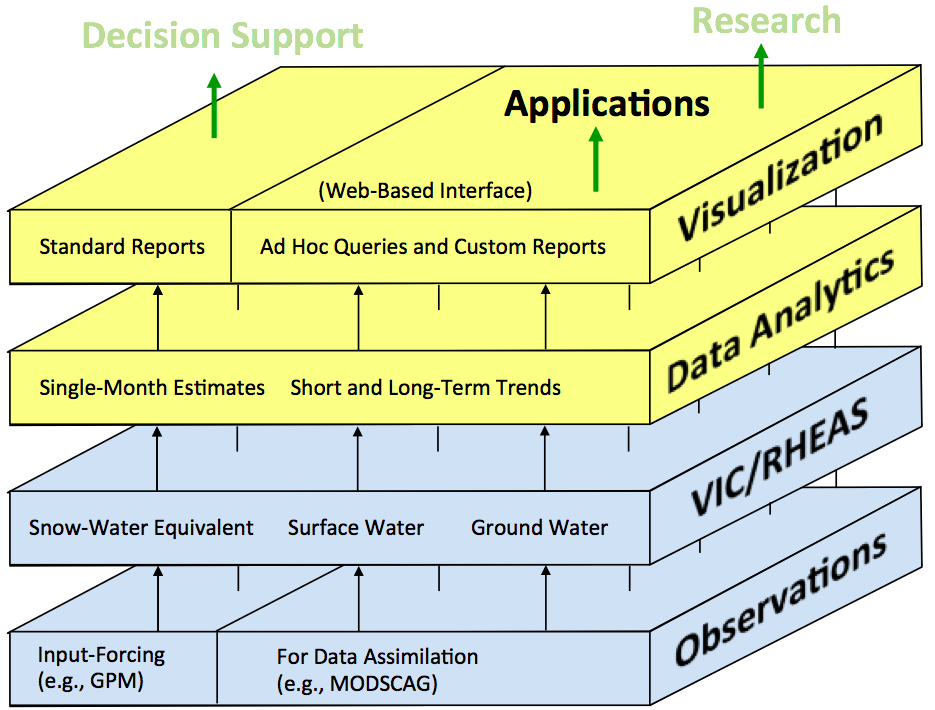} 
 \caption{The workflow layers for WaterTrek. From raw observations of various forms of water short and long term trends are derived. These are then visualized and fed to query based systems to answer research questions.}
   \label{fig:wswma}
\end{center}
\end{figure}

\subsection{Ontologies}
\label{subsec:ontology}

Metadata are data about data and can be used to connect different datasets. An ontology is the methodical description of the semantic interrelationship between metadata elements. To make the entire process of data fusion from diverse sources more standardized, one needs good ontologies.  

Earth science does have its own ontologies and related communities e.g. ONTOLOG (http://ontologforum.org), OGC, Semantic Web for Earth and Environment Terminology (SWEET) (\cite{Raskin2003}) etc. Too many distinct ontologies defeat the purpose though - it is like having multiple standards that do not interface well (\cite{Uschold2004}). Standardization in ontologies is thus a fundamental requirement. That is possible by starting from a core ontology and extending it in areas that interface with it, creating mappings, and gradually encompassing the entire discipline.

That is what is being done by using the framework for the Planetary Data System (PDS) ontology (\cite{Hughes2016}). Built into the PDS ontology is interoperability at multiple levels: (1) Agency to agency - here it is important that the same standards have been adopted by both the agencies. The common terminology and skeletal standards then allow bringing in additional agencies seamlessly, (2) Semantic level - this is governed by commonality in concept definitions. Programmatic interoperability also becomes possible through the use of such common namespaces, (3) Application level - semantic support at the systems interaction level allows users to treat seemingly different applications on par and use them transparently with each other. The resulting data model, PDS4 (\cite{Hughes2009}), captures the knowledge about the planetary science digital repository at several levels of specificity and provides a means by which both humans and machines can “communicate” about the digital content of the repository. The interoperability is enabled by multi-level governance at the discipline, mission etc. levels. A standard vocabulary then results.

The Zachman Framework of Enterprise Architecture (\cite{Zachman1987}) partitions elements into ``why'', ``how'', ``what'', ``who'', ``where'' and ``when''. The PDS4 Information Model encompasses the ``what'' element i.e. the data being processed or archived. By being agnostic to the ``how'' element, it stays flexible to incorporate a variety of ``what''s. The VOEvent/Skyalert framework (\cite{Williams2012}) used for astronomy transient alerts use the same terminology, but also incorporate the ``where'' and ``when'' as the communication is primarily about events, generally with real-time requirements. PDS4 thus includes aspects of interoperability in its information model, and in the design and implementation of the infrastructure supporting the archive holdings. The astronomy/PDS ontology models are being applied to bioinformatics as well as geoinformatics.

\section{Hydrology}
\label{sec:hydrology}

Western States Water Mission (WSWM) has developed WaterTrek, an interactive web-based analytics environment.
It takes in data with multiple resolutions and provides timely actionable information. This involves a close collaboration of hydrological modeling and data science expertise in a mission-style project architecture.

It is critical, in the big data era, that such a framework support different architectural decisions that may change over time (\cite{Rutledge2014}). Thus, a higher level concept of data management and analysis needs to be built. The framework consists of a reusable software stack with primary components like access, computation, workflow, storage, etc. separated. Plug-in algorithms can then support data fusion, classification, and so on. 

Earth data are typically in the form of vectors and areas. Analytics allow averaging over time-periods as well as area/polygons. Data-fusion in this context allows (1) populating the datasets with metadata as needed, (2) ensuring compatibility of metadata through ontology mapping, and (3) constructing interpolated surfaces and data-cubes using time-series data from the suite of water storage observations. For diverse systems, metadata homogenization is required as also query-mapping.

The geographic distribution of the large datasets involved imply that the entire datasets can not be moved on demand for computation. Fig.~\ref{fig:wswmarch} shows the architecture adopted in such a scenario. Data are organized in a data store with indices to allow scalable access. Measurements are stored in Apache Accumulo, a NoSQL DB, which combined with geospatial indexing allows for the handling of feature and raster data by GeoMesa (\cite{Fox2013}) and GeoTrellis (http://geotrellis.io/) respectively. Raw data are backed up using HDFS. Spark framework is appropriate for the scalable analytics required as well as for analytic pipelines. Analytics include computing various short- and long-term trends, in order to answer questions about inter-relationsships, and predicting at least the near future (see Fig.~\ref{fig:wswma}).

\section{EarthCube}
\label{sec:earthcube}

EarthCube (https://www.earthcube.org/) is a NSF funded initiative to transform Earthscience by developing cyberinfrastructure for sharing, visualizing, analysing Earth data and resources. The funding started in 2011 and is expected to continue at least until 2022 in the current round. It is a community driven effort to build interoperability standards, better integration of existing and newer datasets. A side-effect will be the democratizing of data which is in line with NSF requirements of data release and reuse.

\begin{figure}[t]
\begin{center}
 \includegraphics[width=5.0in]{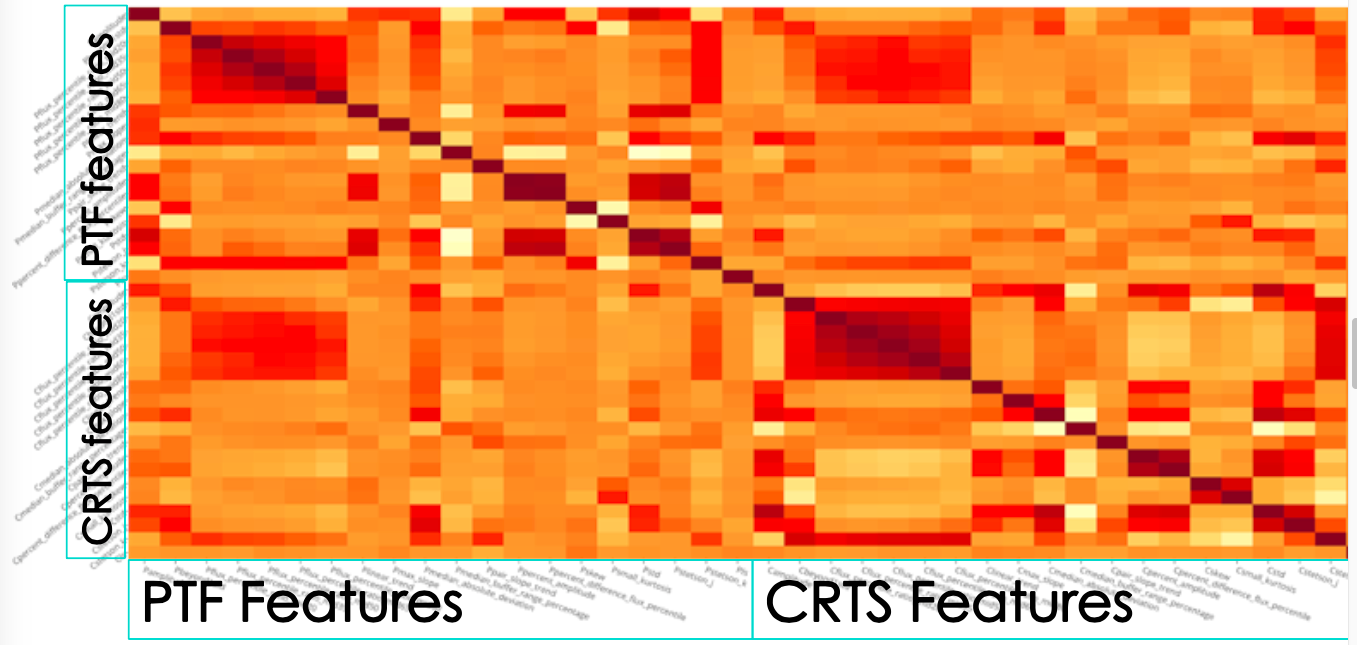} 
 \caption{Correlations for a large number of statistical features for time-series from CRTS and PTF datasets for a set of about 50000 variable sources. Top-left and bottom-right quadrants are self-correlations for the two surveys. The other two quadrants compare the two surveys with each other. It is clear that correlation between identical statistical abstractions from time-series from two different surveys is almost non-existent. Fig. courtesy J. Li, S. Vaijanapurkar.}
   \label{fig:crosscor}
\end{center}
\end{figure}

Many projects - serving as building blocks - are being funded under EarthCube. Examples include (a) BCube (https://www.earthcube.org/group/bcube) to explore the use of brokering technologies to make data discovery, sharing and access easier, (b) Scalable Communiy-Driven  Architecture (SC-DA - https://www.earthcube.org/group/scalable-community-driven-architecture), (c) ECITE (https://www.earthcube.org/group/earthcube-integration-testing-environment-ecite) an integration and test environment creating a scalable, distributed computing facility for EarthCube as well as community driven demonstrations. There are several  other building blocks starting to provide niche services. Joint Initiative on Data Science and Technology established between JPL and Caltech is involved in (b) and (c) above and drawing heavily on their experience from astronomy, Planetary Data System, and Earth. 

For example, for SC-DA (in which Element84 is also involved) we are developing a conceptual architecture that aims to serve as the blueprint for the definition, construction, and deployment of both existing and new software components to ensure that they can be unified and integrated into an evolutionary national geoinformatics infrastructure for data that is part of EarthCube. It includes definition of process, technical, and information architectures based on a set of guiding architectural principles that examine the lifecycle of data to ensure that different stakeholder needs are addressed.  The effort will be to develop a scalable and extensible approach that can support both a data and computationally intensive environment to enable scientific data management and discovery.

Similarly, in ECITE we are actively engaging EarthCube and the wider geoinformatics community in the definition of requirements, design, and testing of the seamless federated system consisting of scalable and location independent distributed computational resources (nodes) across the US. The hybrid federated system will provide a robust set of distributed resources utilizing public and private cloud capabilities. Resources from four institutions viz. George Mason University (GMU), Amazon Web Services (AWS), Xtreme Science and Engineering Data Environment (XSEDE), and the California Institute of Technology (Caltech) have been integrated in the first steps for the DC2 deployment (http://ecite.dc2.gmu.edu/dc2us2/).

\begin{figure}[t]
\begin{center}
 \includegraphics[width=5.0in]{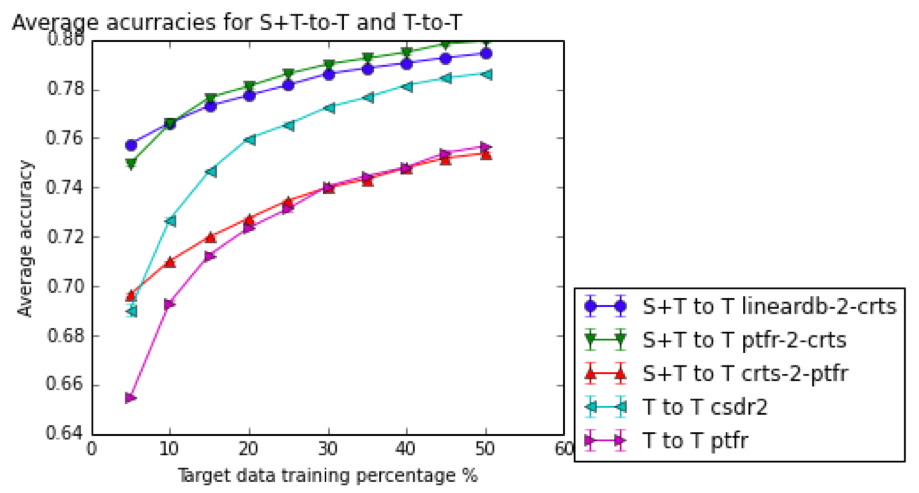} 
 \caption{Using data from CRTS DR2, LINEAR, and PTF r-band surveys, classification percentages are shown when adding an increasing fraction of examples from the target domain to the training set otherwise made from the source domain. The improving classification indicates that one can start useful classification even before the sample from the target survey is large enough to create its own training set. Fig. courtesy J. Li, S. Vaijanapurkar.}
   \label{fig:coda}
\end{center}
\end{figure}

\section{Domain Adaptation}
\label{sec:da}

As an example of a novel methodology that may be useful in different scientific areas, we consider Domain Adaptation.  We illustrate it with an example from astronomy, but it can be easily generalized to other fields. Domain Adaptation allows us to combine datasets with different resolutions, time-spans, time-spacing, as well as wavelength ranges. It is thus an ideal tool for the kind of datasets described earlier. In general when we obtain time-series data for a sample of objects using two such surveys, and then a set of statistical abstractions from the time series, they do not correlate well due to the differences mentioned above. Thus when we used data from CRTS, the Palomar Transient Factory (PTF; http://www.ptf.caltech.edu/), and the Lincoln Near-Earth Asteroid Research (LINEAR; http://neo.jpl.nasa.gov/programs/linear.html) survey data the diagonals of the heatmap representing the cross-survey correlations confirmed that (see Fig.~\ref{fig:crosscor} for the CRTS-PTF correlations, or the lack thereof). We used 50,000 variable objects from CRTS in this comparison (\cite{Drake2014}).

The lack of correlation due to differing properties does not mean that the datasets can not be combined, but just that cleverer ways are required to make full use of them. Domain Adaptation comes in multiple flavors. With the astronomy datasets we have had success using CO-Domain Adaptation (CODA) where one adds different fractions of the target sample set to the source sample set in order to improve classification for the target set (see Fig.~\ref{fig:coda}).

In exactly the same fashion we will be applying domain adaptation to earth science datasets with varied resolutions, and temporal coverages to acquire a holistic overview of a multitude of observables e.g. those related to water.

\section{Discussion}
\label{sec:discussion}

Methodology transfer can almost never be unidirectional. Diverse fields grow by learning tricks employed by other disciplines. The important thing is to abstract data - described by meaningful metadata - and the metadata in turn connected by a good ontology. We have described here a few techniques from astroinformatics that are finding use in geoinformatics. There would be many from earth science that space science would do well to emulate. Even other disciplines like bioinformatics provide ample opportunities for methodology transfer and collaboration. With growing data volumes, and more importantly the increasing complexity, data science is our only refuge. Collaboration in data science will be beneficial to all sciences. 

We end this brief with a description of an Augmented Reality (AR) based outreach tool created in the image of Pok\'emon GO. In Pok\'emon GO, the mobile-based AR game that caught the world by storm a few months ago, as you traverse the surface of the earth, magical creatures pop-up in the mobile-map of your surroundings. Such geolocation techniques are used extensively by many other tools including various traffic apps, and those locating different services around you. Sky maps, analogous to earth maps, exist and in these can be placed rapidly fading transient sources found by surveys (e.g. CRTS today and the Large Synoptic Survey Telescope (LSST; http://www.lsstcorp.org) in the near future).  Instead of being at a specific location, one just points the mobile device to different parts of the visible sky to ``catch'' transients. That is an easy way to disseminate information about such events to the public at large (outreach), but it can also be used as a citizen science tool with contributions to classification of sources.  We had developed a tool to push CRTS events to mobile devices and now with undergrads at SUNY Oswego we are working on an AR based system with gamification elements for astronomical transient sources. Different people can choose to look for specific types of transients just as one would subscribe to specific streams from a brokering service. Additional information on each source is then retrievable, including light-curves, archival observations at various wavelengths and times, positional and other ancillary information. This complex data ensemble can trigger useful high-level connections in human neural networks that can be fed back into the system for improved classification and iterative machine learning.
Thus methodology transfer doesn't need to even be between sciences.  There are opportunities for many industries to learn from one another.  Citizen science can certainly learn from other industries that are reaching massive groups of people through mobile technologies, AR, and gaming. Here too, a clear ontology-aided workflow is important, a mantra worth repeating. 

\section*{Acknowledgements:} We acknowledge partial support from the NSF grants ICER-1343661 (AAM, DC, EL, JSH, SGD), ICER-1541049 (AAM, EL, SGD), AST-1413600 (AAM, SGD), AST-1518308 (AAM, SGD), and ACI-1640818 (AAM, DC, JSH, SGD).

\end{document}